\documentclass[12pt]{iopart}

\usepackage{tikz}
\usetikzlibrary{mindmap}
\usepackage{hyperref}
\usepackage{subfigure}
\usepackage{makecell} 
\usepackage{xcolor}

\definecolor{darkgreen}{rgb}{0,0.5,0}
\begin{document}

\title{Artificial Intelligence in Glioma Imaging: Challenges and Advances}
\author{Weina Jin$^1$, Mostafa Fatehi$^2$, Kumar Abhishek$^1$, \\
Mayur Mallya$^1$, Brian Toyota$^3$, and Ghassan Hamarneh$^1$}

\address{$^1$School of Computing Science, Simon Fraser University, Burnaby, Canada}
\address{$^2$Division of Neurosurgery, The University of British Columbia, Vancouver, Canada}
\address{$^3$Department of Surgery, Kingston General Hospital, Queen’s University, Kingston, Canada}
\ead{\mailto{weinaj@sfu.ca}, \mailto{fatehi@alumni.ubc.ca}, \mailto{kabhishe@sfu.ca},\\ \mailto{mmallya@sfu.ca},
\mailto{hamarneh@sfu.ca}}
\vspace{12pt}

\begin{abstract}

Primary brain tumors including gliomas continue to pose significant management challenges to clinicians. While the presentation, the pathology, and the clinical course of these lesions are variable, the initial investigations are usually similar. Patients who are suspected to have a brain tumor will be assessed with computed tomography (CT) and magnetic resonance imaging (MRI). The imaging findings are used by neurosurgeons to determine the feasibility of surgical resection and plan such an undertaking. Imaging studies are also an indispensable tool in tracking tumor progression or its response to treatment. As these imaging studies are non-invasive, relatively cheap and accessible to patients, there have been many efforts over the past two decades to increase the amount of clinically-relevant information that can be extracted from brain imaging. 
Most recently, artificial intelligence (AI) techniques have been employed to segment and characterize brain tumors, as well as to detect progression or treatment-response. However, the clinical utility of such endeavours remains limited due to challenges in data collection and annotation, model training, and the reliability of AI-generated information.

We provide a review of recent advances in addressing the above challenges. First, to overcome the challenge of data paucity, different image imputation and synthesis techniques along with annotation collection efforts are summarized. Next, various training strategies are presented to meet multiple desiderata, such as model performance, generalization ability, data privacy protection, and learning with sparse annotations. Finally, standardized performance evaluation and model interpretability methods have been reviewed. We believe that these technical approaches will facilitate the development of a fully-functional AI tool in the clinical care of patients with gliomas.

\end{abstract}

%
\noindent{\it Keywords}: Glioma imaging, Brain radiomics, Machine learning, Deep learning
%
%
\maketitle
%
%

\section{Introduction}
\begin{figure}
\centering
\includegraphics[width=0.7\textwidth]{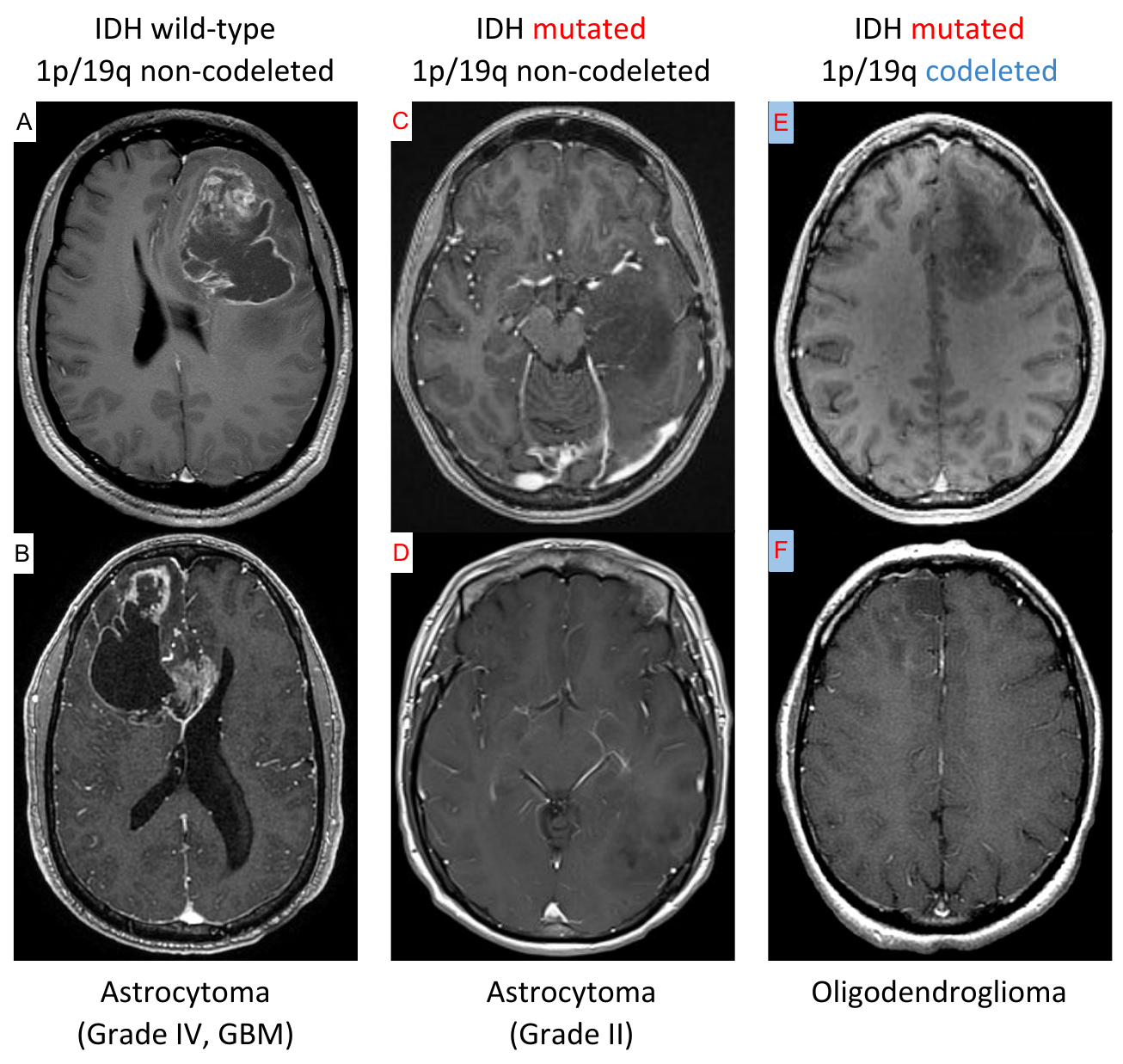}
\caption{There is considerable variability in MR images obtained from patients with gliomas. Panels A and B were obtained from patients with glioblastoma (grade IV astrocytoma). Panels C and D from patients with a low grade astrocytoma (IDH mutated). Panels E and F from patients with oligodendroglioma (IDH mutated and 1p/19q codeleted). All images are axial T1-weighted with gadolinium.}
\label{fig:mri}
\end{figure}

Gliomas are the most common primary tumor of the brain and spine, representing 80\% of malignant brain tumors. These lesions arise from astrocytes, oligodendrocytes or ependymal cells and cause significant morbidity and mortality. The most aggressive subtype,  glioblastoma accounts for $\sim$45\% of all gliomas and affected patients have a 5-year overall survival of $\sim$5\%~\cite{Ostrom2014}. These tumors exhibit an incredible degree of inter and intra-tumoral heterogeneity which leads to variable disease presentation and prognosis (see Figure \ref{fig:mri} as an example). This variability was considered by Bailey and Cushing when they initially described these lesions ninety years ago~\cite{Bailey1929}. The advent of molecular genetics and improvements in techniques like immunohistochemistry and DNA sequencing have led to novel diagnosis classification systems that more accurately predict disease progression and overall survival. However, the personalization of treatment with targeted therapies remains an ongoing challenge for most gliomas.   

Astrocytomas and oligodendrogliomas constitute the great majority of brain gliomas in adults~\cite{Eckel-Passow2015, Louis2016}. These tumors were classically described and graded based upon histopathological features. However, these diagnostic criteria led to significant inter-observer differences and did not correspond to prognosis very well~\cite{Weller2015}. Seminal studies by Cairncross et al. highlighted the importance of chromosome 1p/19q co-deletion for diagnosing oligodendrogliomas~\cite{Cairncross2008}. Later studies found that diffuse gliomas with a mutation in the isocitrate dehydrogenase (IDH, a key rate-limiting enzyme in the energy metabolism cycle~\cite{pmid31263678}) gene had a much better overall survival than unmutated cases (IDH wild-type) ~\cite{Sun2013, Tateishi2017}. These findings led scientists to incorporate genetic characteristics in the new classification of gliomas in 2016~\cite{Eckel-Passow2015, Louis2016}.

The recent primacy of molecular genetics in diagnosing gliomas has coincided with improving knowledge of the aberrant pathways involved in disease progression and susceptibility to chemotherapy and radiation therapy. Consequently, clinicians have attempted to protocolize the treatment of low- and high-grade gliomas. Stupp et al. found that concurrent radiation and temozolamide (TMZ), an alkylating chemotherapeutic agent, improved overall survival in patients with glioblastomas~\cite{Stupp2005}. Moreover, the greatest benefits from TMZ are seen in patients with methylguanine methyl transferase (MGMT) promoter methylation~\cite{Hegi2005, Jha2010}. Adjuvant chemotherapy and radiation are also beneficial for some patients with low grade gliomas. The European RTOG-9802 trial showed that chemotherapy (procarbazine, lomustine and vincristine) after radiation increased progression free survival (PFS) and survival~\cite{Buckner2016}. 

In addition to studying glioma genetics which focuses on specific gene mutations, there is growing interest in understanding these lesions at higher levels. Specifically, genomic studies aim to understand the role of the whole genome on developing gliomas while proteomic studies have investigated the complex interaction of various proteins (normal and aberrant) in affected cells~\cite{Dunn2013}. These fields produce large quantities of data, and hence, rely upon novel computational methods to analyze and confirm findings. As mentioned, brain tumor imaging techniques are non-invasive and  relatively cheap. Thus, they are integral in diagnosis, surgical planning and prognostication. The prospect of obtaining genomic information from  brain imaging led to the establishment of a new field: radiomics~\cite{Gutman2013, Macyszyn2016, Zinn2017}. 

\begin{figure}
\centering
\includegraphics[width=0.75\textwidth]{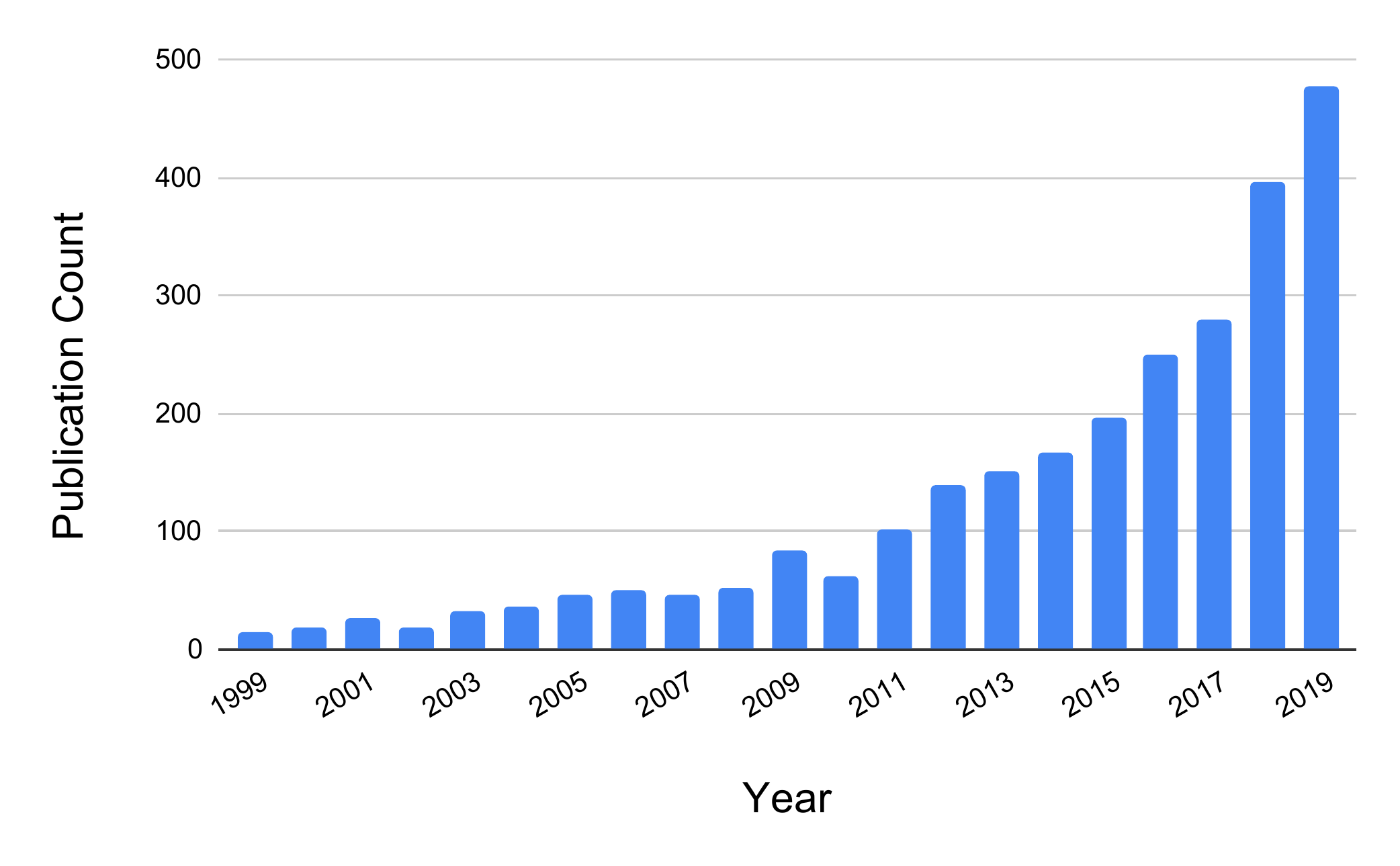}
\caption{Growth of publications on glioma imaging and AI in the past 20 years by PubMed searching (the search keywords can be found at: \href{https://bit.ly/2KgeDNx}{\url{https://bit.ly/AI_in_glioma_imaging}}).}\label{fig:growth}
\end{figure}

Glioma radiomics aim at extracting clinically relevant features from glioma imaging as a non-invasive image biomarker, to provide information for glioma diagnosis, treatment, or prognosis. The earlier studies used statistical methods to identify the significant radiographic features associated with the clinical outcome. In recent years, artificial intelligence (AI), represented by machine learning (ML) and deep learning (DL), has enhanced glioma radiomics research. This is largely because of the increasing volume of brain imaging data, growing computational power, and advances in AI algorithms. Compared to traditional statistical analysis methods, machine and deep learning are able to leverage larger amounts of clinical data with high dimensionality, yet they also require more training data, longer training time and more computational resources, and sacrifice interpretability (and in turn trustworthiness) for predictive power. DL is a subset of ML which is more prominent for the above problems, but it has the advantage of automatically learning features from raw input data, instead of manually engineering features as in ML. Because of the superior predictive power, ML and DL have been widely applied to a variety of glioma imaging-related tasks including: (1) \textbf{brain tumor segmentation}: quantifying tumor volume or segmenting the tumor region for downstream analysis tasks, and (2) \textbf{predictive tasks (classification or regression)}: identifying the tumor type (e.g., distinguishing oligodendrogliomas from astrocytomas), grade, molecular subtypes, or genetic mutation~\cite{Zinn2017, Rathore2019}, and predicting patients' treatment response, length of survival, prognosis, or recurrence (e.g., differentiating glioma recurrence from radiation necrosis and pseudo-progression~\cite{Rathore2019}). These advances have been summarized in previous surveys on DL in glioma imaging~\cite{Shaver2019, Korfiatis2019, Rudie2019}, ML in glioma imaging applications~\cite{Villanueva-Meyer2018, Lotan2019}, brain tumor radiomics~\cite{Zhou2018d, Booth2019} and neuroimaging biomarkers~\cite{Woo2017}.

The introduction of AI to neuro-oncology imaging has been met with great enthusiasm from clinicians. This is partially evidenced by the exponential growth in the number of articles reporting findings. Figure \ref{fig:growth} shows the increasing number of publications involving ML/DL on glioma imaging in the past two decades. Despite significant breakthroughs in this field, AI has not yet been applied to the clinical care of patients with gliomas, due to practical challenges in data collection, model training, and the lack of trust in performance and generalizability. To bridge the clinical utility gap in applying AI to neuro-oncology, our review focuses on recent technical advances addressing various challenges.

\section{Challenges and technical approaches of applying AI in glioma imaging}
\begin{figure}
\center{\includegraphics[width=\textwidth]{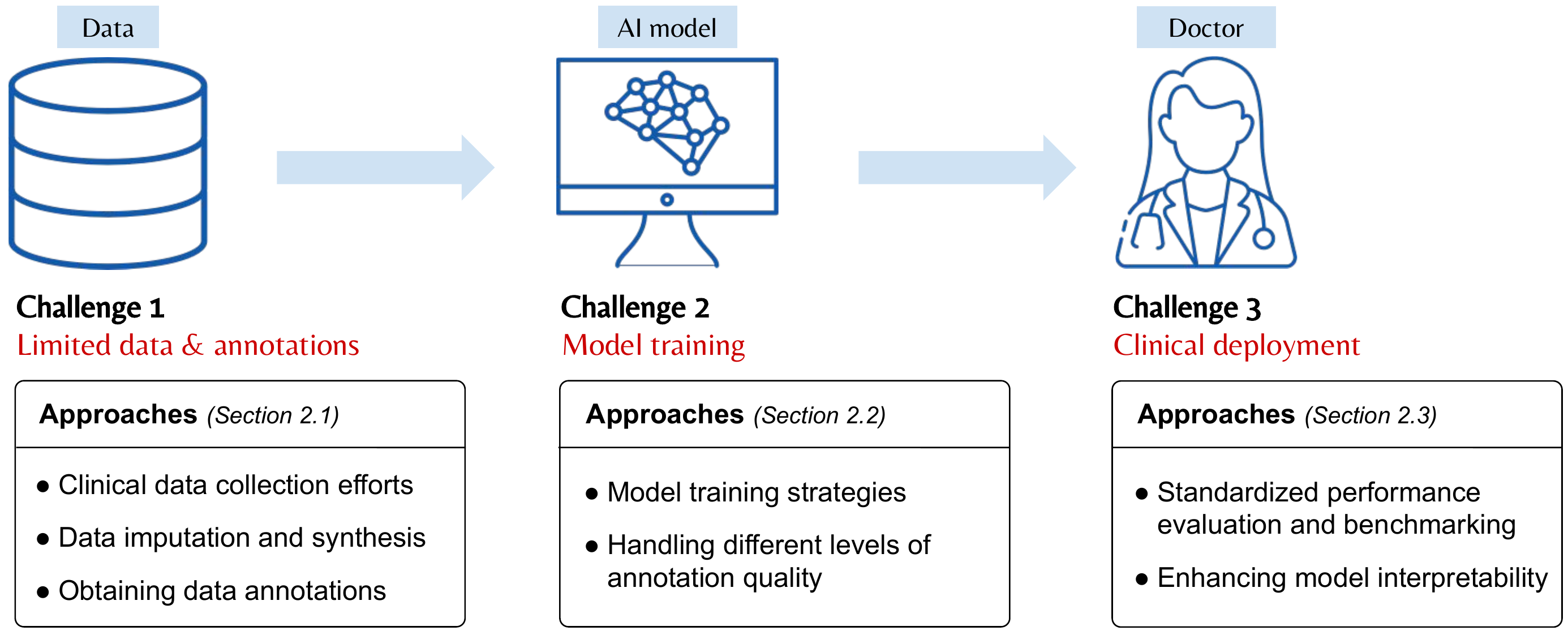}}\caption{The three major challenges and technical approaches to applying AI in glioma imaging.}
\label{fig:overview} 
\end{figure}

Although machine and deep learning techniques have exhibited great potential in analyzing glioma images, their implementation in clinical care remains an elusive goal. Several recent reviews have broadly summarized the challenges in applying AI to clinical medicine (some of which also apply to neuro-oncology)~\cite{He2019, Gilvary2019, Shah2019}. These challenges involve the full life-cycle of developing an AI model, from \textbf{1) obtaining the training data}, to \textbf{2) training the AI models}, to \textbf{3) evaluating and deploying the AI model to clinical settings}. Non-technical challenges also pose practical constrains in developing AI techniques, including: patient data safety and privacy issue; ethical, legal and financial barriers to developing and distributing tools that may impact a patient's treatment course; medical authority regulation, usability evaluation, clinical acceptance, and medical education around the implementation of the AI-assistive tools. Next, we review the recent advances in glioma imaging research addressing the three main challenges. Figure \ref{fig:overview} illustrates the relationship of the challenges and their corresponding approaches. 

\subsection{Obtaining data and annotations}     
\begin{figure}
\centering
\includegraphics[width=\textwidth]{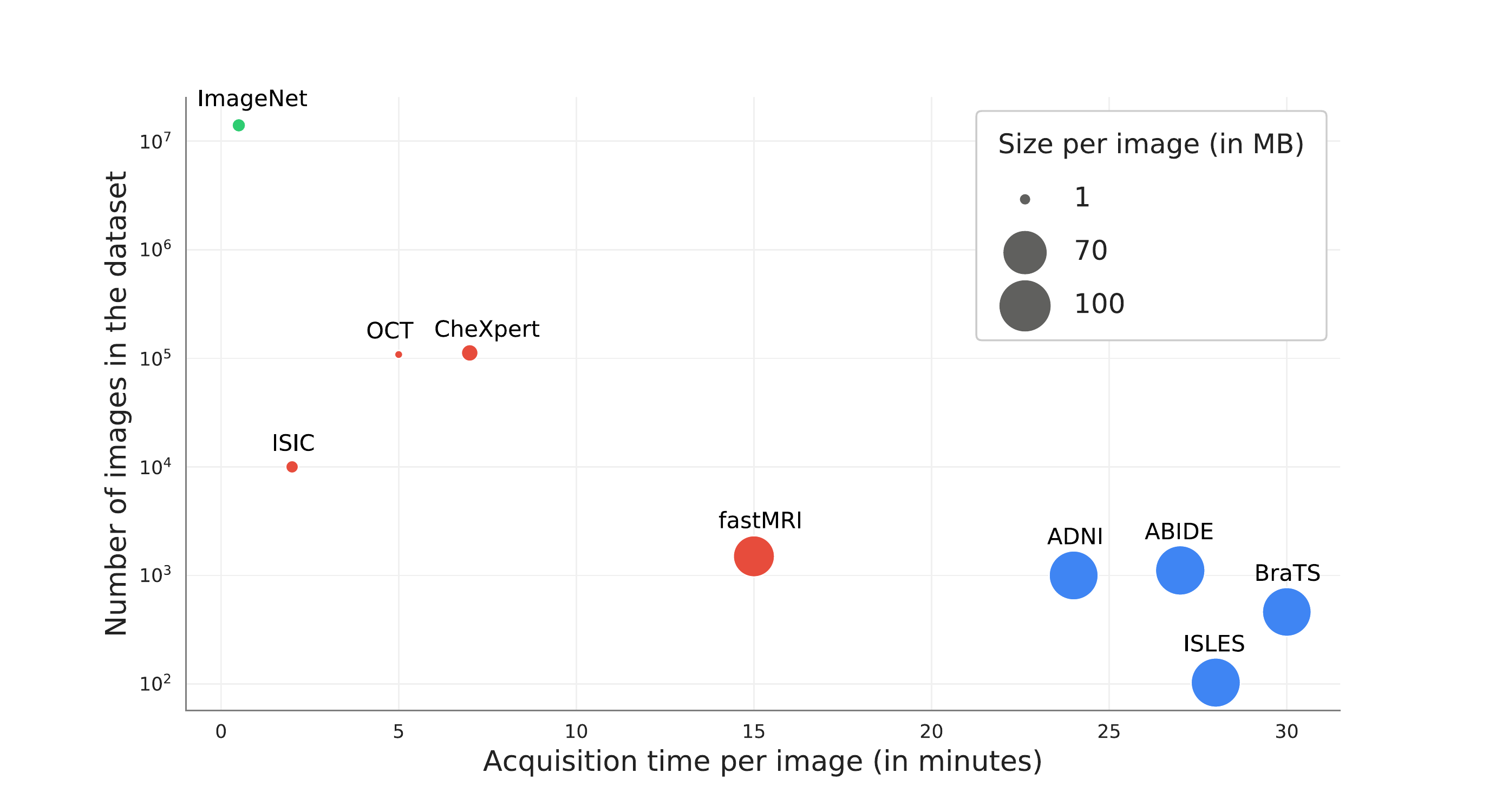}
\caption{The comparison of natural, medical and brain image datasets with respect to their dataset size (y-axis), the image acquisition time (x-axis), and the size per image (shown by the dot size). 
\textbf{Brain} imaging datasets (\textcolor{blue}{blue dots in the figure}): ADNI (Alzheimer's Disease Neuroimaging Initiative)~\cite{adni}, ABIDE (Autism Brain Imaging Data Exchange)~\cite{DiMartino2013,DiMartino2017}, BraTS (Brain Tumor Segmentation)~\cite{menze2014multimodal}, and ISLES (Ischemic Stroke Lesion Segmentation)~\cite{Maier2017}. Other \textbf{medical} imaging datasets (\textcolor{red}{red dots}): OCT (retinal optical coherence tomography images)~\cite{kermany2018labeled}, CheXpert (chest radiographs)~\cite{Irvin2019}, ISIC (International Skin Image Collaboration Melanoma Project)~\cite{isic}, and fastMRI (knee MRI)~\cite{DBLP:journals/corr/abs-1811-08839}. \textbf{Natural} image dataset (\textcolor{darkgreen}{green dot}): ImageNet~\cite{deng2009imagenet}.}
\label{fig:dataset}
\end{figure}

Machine and deep learning models require large amounts of training data to learn from, so that their performances can be generalizable in real-world applications. One prominent example is the ImageNet dataset (containing over 14 million images of natural scenes)~\cite{deng2009imagenet} that has greatly advanced AI analysis of images. However, the field of glioma imaging is lacking such comparable large-scale, consolidated public datasets.  The prevalence of gliomas is relatively low; which, combined with the acquisition cost and time, hinders to the collection of large datasets. Additionally, medical data is protected by patient privacy and safety laws, making it difficult to share these images. Furthermore, the cost of data labelling can be quite high since it requires annotation by medical experts~\cite{Woo2017}. Figure \ref{fig:dataset} shows a comparison of several popular natural, medical and brain imaging datasets with respect to these image acquisition factors. As shown in the figure, natural images (ImageNet) and certain medical imaging modalities, such as dermoscopic images (ISIC) are quick to acquire and generate small file sizes, whereas brain imaging datasets take longer to acquire and are large in size.

Several strategies have been employed to meet the challenge of limited data and annotations: (i) data sharing platforms and initiatives to build a collective glioma imaging database; (ii) data imputation and synthesis techniques to address issues of missing data and lack of sufficient data, and (iii) active learning and crowd-sourcing to facilitate data annotation. Next, we discuss recent advances in these approaches. 

\subsubsection{Clinical data collection efforts\\}

International and multi-institutional efforts have been made in the past decades towards collecting and sharing glioma imaging data. 
The TCIA/TCGA (The Cancer Imaging Archive/The Cancer Genome Atlas) service provides large publicly-available datasets, and is a part of a collective effort by the National Institutes of Health (NIH), the National Human Genome Research Institute (NHGRI), and multiple hospitals to encourage the sharing of clinical, imaging (including radiographic and pathological images) and genomic datasets for cancer research~\cite{Clark2013}. There are about 1,200 raw brain images related to glioma in the TCIA dataset. The Brain Tumor Segmentation (BraTS) challenge dataset~\cite{menze2014multimodal} derived from TCIA contains 461 expert-assessed scans (including low- and high-grade gliomas) from 19 institutions. 
As the largest publicly-available glioma imaging dataset, BraTS is critical for benchmarking and has helped advance AI applications in glioma image analysis.


\subsubsection{Data imputation and synthesis methods\\*}
As mentioned, data heterogeneity and data  paucity are two prominent issues in the clinical adoption of AI in glioma imaging. As such, data imputation approaches have been proposed to handle scenarios such as missing MRI sequences, missing voxel intensities, low-resolution or artifact-degraded scans. Similarly, data synthesis and augmentation techniques have been proposed to expand the available imaging data.

\begin{itemize}
    \item \textbf{Data imputation methods to deal with missing data}\\
    Several conventional machine learning methods have been proposed for MR image imputation. Cruz et al.~\cite{cruz2019imputation} simulated two types of missing morphological measures in $T_1$-weighted brain MRI scans: missing completely at random (MCAR) and missing not at random (MNAR). They also evaluated four data imputation techniques representative of distinct statistical approaches: substituting the missing value with the mean of all subjects (univariate approach), k-nearest neighbors (inferential approach), low-rank matrix approximation (analytical approach), and random forest (statistical learning approach). The random forest-based approach was shown to be the most accurate in recovering the gold standard results. 
    Jog et al.~\cite{jog2017random} proposed a random forest-based supervised image imputation approach, which performs a non-linear regression in the feature space to predict the intensities in various contrast-enhanced MR sequences. They evaluated their proposed approach by synthesizing 1) $T_2$-weighted images from skull-stripped $T_1$-weighted images 2) FLuid Attenuated Inversion Recovery (FLAIR) images from $T_1$-weighted, $T_2$-weighted, and $P_D$-weighted images and 3) whole head $T_2$-weighted images (non-skull-stripped) from $T_1$-weighted images. Their method demonstrated an improvement in the synthetic image quality over state-of-the-art image synthesis algorithms, and the downstream segmentation task performed similarly for real and imputed images. Dalca et al.~\cite{8444451} leveraged the shared intrinsic local fine scale structural similarity in a collection of medical images. They used a generative model, Gaussian mixture model (GMM), to perform brain MRI restoration in a sparsely sampled scan, and demonstrated superior performance as compared to state-of-the-art up-sampling super-resolution methods. They cautioned, however, that imputed data should not be used in clinical evaluation; instead the brain mask obtained from the restored scans can be applied to the original scans and improve subsequent analyses.

    Recently, deep learning-based approaches have been proposed for data imputation. Chartsias et al.~\cite{chartsias2017multimodal} proposed a multi-input multi-output deep model trained end-to-end, with encoders and decoders for each MRI sequence. The latent representations from multiple encoders were fused, and a combination of loss functions were used to ensure that the multiple latent representations from multiple modalities are similar. 
    Another strategy for MRI imputation relies upon generative adversarial networks (GANs)~\cite{goodfellow2014generative}. 
    Welander et al.~\cite{welander2018generative} compared the performances of two image-to-image translation methods: CycleGAN~\cite{zhu2017unpaired} and UNIT~\cite{liu2017unsupervised} for the generation of $T_1$ and $T_2$ MRI sequences; the generated $T_2$ sequences had higher mean absolute error (MAE) and lower peak signal-to-noise ratio (PSNR) than the generated $T_1$ sequences. Several other GAN-based methods for imputing missing MRI sequences have been proposed~\cite{orbes2018simultaneous, dar2019image, sharma2019missing, yurt2019mustgan, kwon2019generation}, including CollaGAN~\cite{lee2019contrast} which investigated the reconstruction of various sequences. They concluded that while $T_1$, $T_2$, and $T_2$-FLAIR can be imputed from other sequences, contrast-enhanced $T_1$-weighted sequences cannot be efficiently imputed.

    \item \textbf{Data synthesis methods for data augmentation}\\
    Similar to data imputation techniques using GANs, several data synthesis-based augmentation approaches using GANs have been proposed. To synthesize new images, Han et al.~\cite{han2018gan} trained two generative models, a deep convolutional GAN (DCGAN)~\cite{radford2015unsupervised} and a Wasserstein GAN (WGAN)~\cite{arjovsky2017wasserstein}, on the BraTS dataset to synthesize new brain MR images. To evaluate their performance, the authors conducted a `visual Turing test': doctors were asked to differentiate real MRI from the generated images. Images generated by WGAN (classification accuracy ranged from $53\%$ to $64\%$) were mis-classified by doctors more frequently than those generated by DCGAN (classification accuracy ranged from $54\%$ to $70\%$); thus inferring that WGAN produced more visually realistic images. A similar visual assessment was also conducted by Kazuhiro et al.~\cite{kazuhiro2018generative} where five radiologists (including two experienced neuroradiologists) attempted to classify a random selection of 50 real and synthesized images. The classification accuracy for the neuroradiologists were $55\%$ and $30\%$; indicating that many synthesized images were practically indistinguishable from real MRIs.
    Shin et al.~\cite{shin2018medical} used an image-to-image translation framework~\cite{isola2017image} to predict tumor segmentation masks from brain MR scans and to generate brain MR scans given tumor masks. The latter was then used to augment the segmentation training dataset. Using only the synthetic MR images to train the model followed by fine-tuning on $10\%$ of the real images, the segmentation performance was superior to using only the real images. Jin et al.~\cite{jin2019deep} proposed an image-to-image translation framework for synthesizing brain MR images from brain CT images. They leveraged a combination of adversarial loss, dual cycle-consistent loss, and voxel-wise loss strategies to train. Models trained using the paired data (i.e., CT and corresponding MRI scans from the same patient) and the unpaired data (two independent collections of CT and MRI images) together achieved higher PSNR and lower MAE compared to models trained using only either paired or unpaired data.
\end{itemize}

\subsubsection{Obtaining data annotations\\}

Classically, MRI annotation has required the time and expertise of medical experts, which imposes higher costs than for natural images. As such, there has been interest in 1) making the best use of expert input by annotating in an active learning setting, or 2) bypassing physician expertise and asking non-experts to annotate. We will expound upon each of these strategies below.

Active learning (AL) is a possible workaround to training models with limited annotated data. AL is an ML technique that reduces the annotation workload by automatically deciding which representative samples should be annotated in order to train a model as quickly and efficiently as possible. Damangir et al.~\cite{Damangir654127} proposed using a cascade of support vector machines (SVMs) with an AL scheme to segment white matter lesions from brain MR scans (a problem compounded by the dataset being biased towards normal tissue). They observed that the proposed method yielded accurate results while also achieving the highest sensitivity and specificity scores amongst all methods compared. Konyushkova et al.~\cite{konyushkova2015introducing} proposed leveraging the feature and geometric uncertainties from images to increase the efficiency of AL for tumor segmentation from brain MR images. In a later study, they modeled AL as a regression task ~\cite{konyushkova2017learning}, essentially predicting the expected generalizability of a trained classifier when annotating a specific data point. They used features extracted from a random forest regressor trained on synthetic 2D data to perform tumor segmentation from brain MR scans, and obtained superior results as compared to other active learning methods (such as uncertainty sampling~\cite{10.5555/188490.188495} and active learning by learning~\cite{hsu2015active}) with considerably fewer annotated samples.

Another popular approach of obtaining a large number of annotations quickly is to crowd-source the annotation process; a strategy which has led to rapid advances in computer vision ~\cite{Vaughan2018}. The aforementioned ImageNet dataset~\cite{deng2009imagenet} has been annotated using a crowd-sourcing platform (Amazon Mechanical Turks). Although there may be few medical experts within the labelling group, crowd-sourcing has been shown to be effective in creating large quantities of annotated data; and, it is faster and cheaper than annotation by medical experts~\cite{Orting2019}. In neuroscience research, crowd-sourcing and gamification have helped neuroscientists to explore brain networks by identifying neurons and their synaptic connections ~\cite{Roskams2016,Marx2013}. Prior to annotating, the crowd will be trained by learning from a few examples labelled by medical experts. The labelling results will need to be aggregated from multiple workers to ensure the labelling quality. To date, crowd-sourcing tasks in medical imaging have included classification, localization and segmentation of organs or lesions~\cite{Orting2019}. 
Due to the image variability and the requirement of domain knowledge, crowd-sourcing in neuroimaging, especially in glioma research, is still in its infancy, with only a few studies in this area.  For example, 
to analyze diffuse intrinsic pontine glioma, 823 non-expert participants annotated 5,152 brain MR images to segment brain stem or tumor~\cite{Timmermans}. Rajchl et al. recruited 12 non-expert to annotate super-pixels for fetal brain segmentation. The fully convolutional network trained on crowd-sourcing annotations reached similar performance as the one annotated by experts~\cite{Rajchl2016}.
Keshavan et al. used crowd-sourcing to amplify expert-labelled data for quality control of brain MRI images, and a deep learning model trained on the amplified data reached AUC of 0.99~\cite{Keshavan2019}.

\subsection{Training the model}
Successful training requires the model to meet multiple desiderata such as satisfactory performance, generalizability, data privacy protection, and training with sparsely annotated data. 
\subsubsection{Choosing and training models\\}
Some practical challenges are related to model training, and we list the three most important ones: choosing the optimal model, model generalization, and learning under data privacy constraint.

\begin{itemize}
    \item \textbf{Choosing the optimal model\\}
The proliferation of deep learning models has led to a considerable number of layer designs and model architectures, loss functions, and optimizers to choose from when designing a network. With an infinite space of possible computation graphs, a plethora of architectures were proposed (some of the popular model architectures include AlexNet~\cite{krizhevsky2012imagenet}, VGG16 and VGG19~\cite{simonyan2014very}, GoogLeNet~\cite{szegedy2015going}, ResNet~\cite{he2016deep}, DenseNet~\cite{huang2017densely}, inception architectures~\cite{szegedy2016rethinking, szegedy2017inception}, U-Net~\cite{ronneberger2015u}, etc.).  Moreover, even randomly connected graphs~\cite{xie2019exploring}, as well as strategies to perform neural architecture search (NAS)~{\cite{elsken2019neural, wistuba2019survey}} have been developed in order to find the optimal model architecture. We list the model architectures used by various works in glioma imaging in Table~\ref{comparison}.

\item \textbf{Model generalization\\} Models trained on images from one hospital may not be generalizable or perform equally well on new data from another hospital or image scanner, due to domain distribution shift and the confounding information models exploited~\cite{Zech2018}. 
Regularization strategies are utilized to improve model generalization and prevent over-fitting the training dataset. These include dropout~\cite{srivastava2014dropout}, early stopping~\cite{prechelt1998early}, data augmentation~\cite{zhang2016understanding}, or gather sufficient data from different scanners and/or hospitals~\cite{Aslani2019}. To encourage model generalization among multiple scanners, Aslani et al. proposed to use an auxiliary network and corresponding regularization loss term to learn the domain-specific knowledge~\cite{Aslani2019}. This auxiliary network learned to predict the category of the input scanner's domain, thus encouraging the backbone segmentation network to ignore domain-specific information. The experiments on brain lesion segmentation from 56 different scanning sites showed the proposed method had better generalization performance to data from new sites than other baseline networks.
\textbf{Transfer learning}~\cite{Raghu2019} is also a common technique to deploy models at new site where data distribution may be different from the original training set. To achieve equivalent performance on new data, transfer learning uses models pre-trained on datasets from previous hospitals to re-train the model on new data. Ghafoorian et al. conducted experiments on transfer learning on brain lesion segmentation tasks; and demonstrated  that, without transfer learning, the model completely failed on new data with the same follow-up patients and when images were acquired with different acquisition protocols. With transfer learning on a small set of new training examples, the model substantially outperformed a model trained from scratch with the same size of training data~\cite{Ghafoorian2017}. The authors also observed as the new training data becomes available, fine-tuning can be done from the last full-connected layer to the last few convolutional layers.

Transfer learning, however, can suffer from catastrophic forgetting issues, where the knowledge about the old task may not be maintained when adapting parameters to a new dataset or task. To avoid this pitfall and enable \textbf{continual learning}~\cite{Baweja2018},  several approaches were proposed in the realm of brain segmentation. Garderen et al. applied a regularization called elastic weight consolidation (EWC) during transfer learning~\cite{Garderen2018}. It penalizes large changes in model parameters weighted based on their importance to the old dataset. Research on segmenting low- and high-grade gliomas showed that EWC improved performance on the old domain after transfer learning on the new domain. Conversely, it also restricted the adaptation capacity to the new domain. Karani et al. suggested that learning batch normalization parameters for each scanner and sharing the convolutional filters between all scanners addressed the distribution shift among scanners ~\cite{Karani2018}. The experiment showed this strategy can be adapted to new scanners or protocols with only a few ($\approx 4$) labelled images and without degrading performance on the previous scanners. 

\item \textbf{Learning under data privacy constraint\\}
Training models on data from multiple sites can improve model generalization. However, due to data privacy and regulation issues, data is usually retained within host hospital servers, and is not easily shared. Data privacy protection methods can range from data anonymization, obfuscation (modifying data with noise to provide protection, e.g.: differential privacy)~\cite{Abadi2016}, to federated learning (sharing model parameters rather than raw data) and learning over encrypted data~\cite{Phong}.
\textbf{Federated learning} is a private distributed and decentralized machine learning method that trains the shared model locally with private data without exchanging the raw patient data~\cite{Vepakomma2018a,Vepakomma2018b}. Sheller et al. demonstrated federated learning on the clinically-acquired BraTS data for institution-level segmentation tasks~\cite{Sheller2019}. Their study showed that the performance of federated semantic segmentation models (Dice=0.852) was similar to that of models trained by sharing data (Dice=0.862), and outperformed two alternative collaborative learning methods.
\end{itemize}

\subsubsection{Handling different levels of annotation quality\\}
\begin{figure}
\centering
\includegraphics[width=\textwidth]{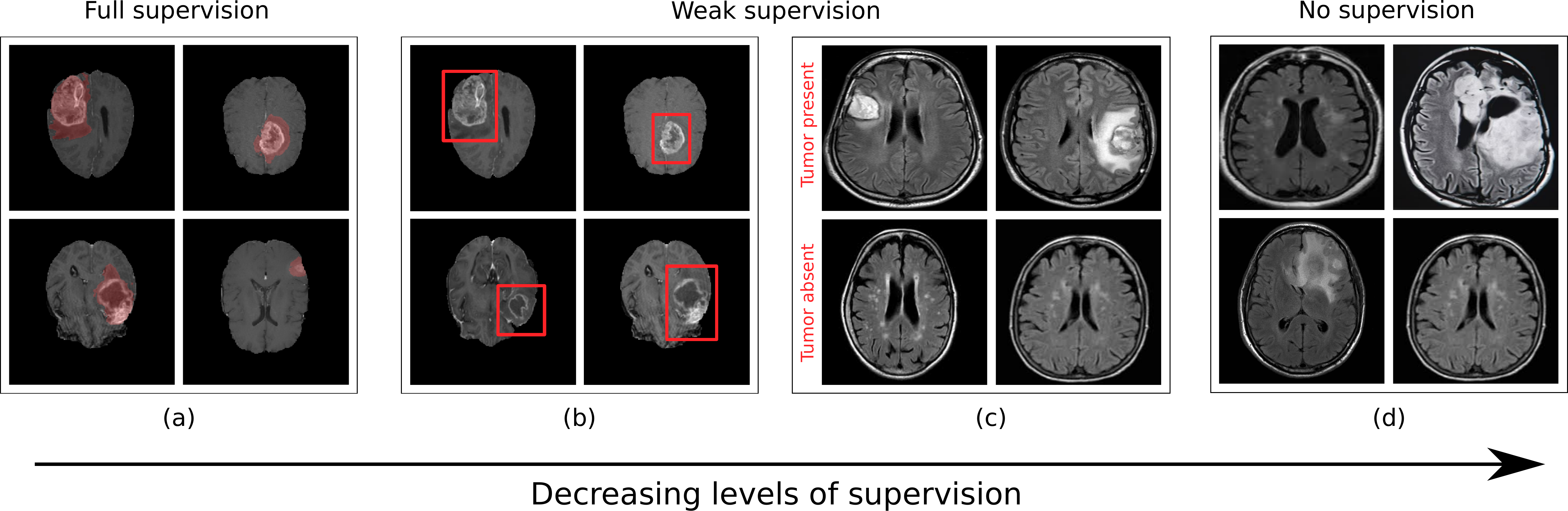}
\caption{Different levels of supervision for a tumor segmentation task. (a) Fully supervised learning uses pixel-level annotations, whereas (d) unsupervised learning does not rely upon labels. Semi-supervised learning uses weak supervision varying from (b) bounding box annotations to (c) just image-level labels.}
\label{supervision_types}
\end{figure}

Machine and deep learning models can be trained with varying levels of supervision. Fully annotated ground truth labels for medical imaging data are expensive and relatively  scarce; however, images with weaker levels of annotation such as bounding box annotations or image level annotations are relatively easier to acquire. Figure~\ref{supervision_types} shows the different levels of supervision for training a tumor segmentation model. A fully supervised system would require manually annotated tumor segmentation regions (pixel-wise labeling). Weaker levels of annotation can vary from bounding box annotations of tumor region or image-level labels indicating whether a tumor is present (semi-supervised learning), or no labels at all (unsupervised learning). For detailed description of semi-supervised medical image analysis techniques, we refer the interested readers to Cheplygina et al.~\cite{cheplygina2019not}.

Conventional machine learning methods have been deployed to learn from sparsely annotated data. Azmi et al.~\cite{azmi2013ensemble} used stationary wavelet transform, edge features, and fractal features in a semi-supervised framework with an ensemble of three algorithms for segmenting brain tissues. Their approach achieved a higher accuracy and precision than fully-supervised approaches, and they noted that supervised methods such as k-nearest neighbor (KNN), SVM, and Bayesian classifiers were not able to perform well with limited data. Blessy et al.~\cite{blessy2015performance} proposed an unsupervised approach for MRI brain tumor segmentation based on optimal fuzzy clustering. They evaluated their approach on MRI scans of 150 patients with low grade gliomas from the BraTS dataset and outperformed four other clustering-based methods. Grande-Barreto et al.~\cite{grande2018unsupervised} used priors about the brain structure and features extracted from the 3D gray level co-occurence matrix (GLCM) for segmenting brain tissue, and observed superior performance when evaluated on synthetic brain MR scans as compared to other state-of-the-art unsupervised approaches.

More recently, there have been several deep learning-based semi-supervised and unsupervised approaches applied to brain MR images. Chen et al.~\cite{chen2018unsupervised} trained variational autoencoder~\cite{kingma2013auto} and adversarial autoencoder~\cite{makhzani2015adversarial} models to learn the latent space representation of healthy brains and introduced a representation consistency constraint in the latent space, leading to a better lesion detection accuracy. Alex et al.~\cite{alex2017semisupervised} proposed using stacked denoising autoencoders (DAE)~\cite{vincent2010stacked} for segmenting high grade gliomas, transfer learning for segmenting low grade gliomas, and a single layer DAE for constructing error maps to improve lesion detection performance. Similarly, Enguehard et al.~\cite{enguehard2019semi} proposed a semi-supervised algorithm for infant brain MRI tissue segmentation based on deep embedded clustering, and demonstrated the robustness of their proposed approach by using the same network, optimizer, and learning rate for two additional natural image classification tasks and achieving superior performance to state-of-the-art semi-supervised methods. Dalca et al.~\cite{dalca2019unsupervised} proposed a probabilistic model and a novel loss function for unsupervised segmentation of brain MRI, and built upon the recent advances in deep learning-based registration methods to jointly estimate the atlas deformation and the image intensity statistics, allowing them to train a deep model on a new dataset without any manually annotated images.

In scenarios with multiple granularities of annotated ground truth, a mixed supervision training approach can be adopted in order to utilize all the available annotations. Zheng et al.~\cite{zheng2018integrating} proposed a multi-atlas label fusion based method to segment the hippocampus in brain MR images by integrating random forest classification models with semi-supervised label propagation. Baur et al.~\cite{baur2019fusing} trained an autoencoder to learn the representation of healthy brain scans, and then used it to detect and segment anomalies in unlabeled scans. A U-Net~\cite{ronneberger2015u} was then trained for supervised segmentation with images paired with their labeled ground-truth data and unseen images paired with the autoencoder output, thereby combining unsupervised and supervised techniques. Mlynarski et al.~\cite{mlynarski2018deep} extended this architecture~\cite{ronneberger2015u}  by jointly performing classification and segmentation, and utilized weakly annotated and fully annotated data to train a deep model for tumor segmentation from brain MRI scans.

\subsection{Clinical deployment}
In this section, we describe approaches which address the main challenges to translating the advances in machine and deep learning into clinical deployment: namely, standardized evaluation metrics and model interpretability.

\begin{table}[]
\centering
\caption{List of deep learning model architectures and evaluation metrics used in glioma imaging.}
\label{comparison}
\resizebox{\textwidth}{!}{%
\begin{tabular}{lll}
\br
Task & Model architecture & Evaluation metrics \\ 
\br
\textbf{Classification}    & VGG16~\cite{Ahmad2019}                  & Accuracy~\cite{Ahmad2019,  Chang1201, Meier2018, Ye2017}                  \\
     & 2D~\cite{Chang1201, Ahmad2019} and 3D ResNet~\cite{Meier2018}                  & Sensitivity or recall~\cite{Ahmad2019, Meier2018, Ye2017}                  \\
     & CNN+LSTM$^{a}$~\cite{Jang2018a}                  & Specificity~\cite{Ye2017}                  \\
     & 3D CNN with gated multimodal unit~\cite{Ye2017}                  & Precision or PPV$^{b}$~\cite{Ahmad2019, Meier2018,  Ye2017, Jang2018a}                  \\
     &                    & F1 score~\cite{Ahmad2019,Meier2018,Jang2018a}                  \\
     &                    & AUC$^{c}$~\cite{Meier2018,Jang2018a}                  \\ \mr
     
\textbf{Segmentation}    & 2D~\cite{orbes2018simultaneous, Karani2018, Sheller2019, mlynarski2018deep} and 3D U-Net~\cite{Aslani2019, Garderen2018}& 
Dice score~\cite{orbes2018simultaneous,Aslani2019,Ghafoorian2017, Garderen2018, Karani2018, Sheller2019, mlynarski2018deep, Chen2018a, Havaei2017}\\
     & Fully convolutional network~\cite{Ghafoorian2017}                  & Specificity~\cite{Havaei2017}                  \\
     & Densely connected CNN~\cite{Chen2018a}                   & Sensitivity~\cite{Aslani2019, Havaei2017}                  \\
     &  TwoPathCNN~\cite{Havaei2017}                  & Precision~\cite{Aslani2019}                  \\
     &                    & False positive rate~\cite{orbes2018simultaneous, Aslani2019}                  \\
     &                    & False negative rate~\cite{orbes2018simultaneous}                  \\ \mr
     
\textbf{Synthesis}    & GAN~\cite{welander2018generative, orbes2018simultaneous, dar2019image, sharma2019missing, yurt2019mustgan, kwon2019generation, lee2019contrast, Huang2019}                  & MSE$^{d}$~\cite{sharma2019missing, Huang2019}                  \\
     &                    & MAE$^{e}$~\cite{welander2018generative, orbes2018simultaneous}                  \\
     &                    & PSNR$^{f}$~\cite{ welander2018generative, orbes2018simultaneous, dar2019image, sharma2019missing, yurt2019mustgan, Huang2019}                  \\
     &                    & SSIM$^{g}$~\cite{dar2019image, sharma2019missing, yurt2019mustgan, kwon2019generation, lee2019contrast, Huang2019}                 \\ \br
\end{tabular}
}\\
\begin{flushleft}
\footnotesize{
$^{a}$LSTM: long short-term memory;
$^{b}$PPV: positive predictive value;
$^{c}$AUC: Area under the receiver operating characteristics; 
$^{d}$MSE: mean squared error;
$^{e}$MAE: mean absolute error;
$^{f}$PSNR: peak signal-to-noise ratio;
$^{g}$SSIM: structural similarity index
}
\end{flushleft}
\end{table}

\subsubsection{Standardized model evaluation\\}
The existing literature reports the performances of AI models using a wide range of evaluating metrics, e.g., accuracy, sensitivity, specificity, area under the curve (AUC), and $F_1$ scores for classification tasks, overlap-based measures such as the Dice score coefficient and the intersection over union measure (also known as the Jaccard index) for segmentation tasks. Table \ref{comparison} summarizes the evaluation metrics used in the literature.
The metrics variability can lead to models being evaluated based on the reported best metric, making horizontal comparisons among the models a difficult task. It is therefore necessary to have a standardized set of evaluation metrics to provide a uniform platform for comparison of different machine and deep learning models. 
One possible approach to deal with this would be to report the model performance using all the conventional evaluation metrics, thereby allowing for assessing the performance of newer models with respect to the established benchmarks~\cite{Luo2016}.

Moreover, high values for a reported  metric of a model do not necessarily translate into superior clinical performance, mainly due to the fact that individual metrics don't capture all the desirable properties for a model~\cite{kelly2019key}. For example, a viable model needs to be accurate, fast, and relatively small; invariably, there are trade-offs between model performance, generalizability, and interpretability.

\subsubsection{Model interpretability\\}
Compared to simple statistical models such as linear models, most machine and deep learning approaches suffer from the notorious problem of using opaque models (black-box models) rather than transparent ones (white-box models). Even though the model parameters and architecture are known, tracing the relationship between input-output pairs is a  difficult challenge. This is partially due to the large number of model parameters; deep neural networks usually have millions of parameters (for example, the popular VGG16 architecture has over 138 million parameters). Furthermore, there are complex representations in high-dimensional space, and multi-layered non-linear mappings from input space to output predictions. In machine and deep learning literature, this challenge is referred to as the model interpretability or explainable AI (XAI) problem, i.e., to open the black box models and reveal how the model makes the predictions in terms that human users can understand~\cite{Doshi-Velez2017}.

Model interpretability is especially important in deploying AI techniques in clinical settings. With a black-box model, the clinical users will only receive a prediction without an explanation or justification. Thus, it is difficult for clinical users to trust such models, adopt the model results as part of the formal and legal medical report, and learn from models to improve their expertise. XAI is therefore regarded as one of the ``missing pieces'' of AI in medicine~\cite{Gilvary2019, Gordon2019}. Although XAI is still an emerging field, there have been some recent attempts to explain AI models on glioma imaging predictive tasks. In the subsequent sections, we introduce different approaches to explaining AI decisions to clinical users, and summarize progress in applying XAI to glioma imaging tasks.

Explanations of the model's prediction can be produced at different levels and using different approaches.
\textbf{Local} and \textbf{global explanations} are explanations at different levels of granularity~\cite{Holzinger2017}. Local explanations explain how the model makes the decision for one data point, usually when doctors are seeking explanations for a particular clinical case, and global explanations explain how the model makes decisions in general. 
Similarly, \textbf{post-hoc} and \textbf{intrinsic (ante-hoc) explanations} are two distinct approaches to generate explanations~\cite{ahmad2018interpretable}. Post-hoc explanation builds a proxy explanatory model to explain the original black-box model. It is especially suitable in explaining black-box models that are already put in use or those that use a specific architecture. In contrast, for models that have not been put into practice or do not have their architecture specified, we can build intrinsic (ante-hoc) explainable models that both perform predictive tasks and explain the predictions.

A prerequisite of useful XAI is that the generated explanations need to be easily understandable by clinical users. Creating explanations for clinical end-users is more challenging since they usually do not have prior knowledge in ML or AI models. 
Jin et al. surveyed the existing XAI literature and identified three explanatory forms that are accessible to non-technical  users~\cite{Jin2019}:
\begin{itemize}
    \item Explaining using \textbf{feature attributes} to identify the most important features that have the highest influence on the predicted outcome.
    \item Explaining using \textbf{examples} to show users instances that the model regards to be similar to the query image, or are typical to the prediction.
    \item Explaining using \textbf{decision rules} to show users the decision flows that the model follows. 
\end{itemize}
All the features, examples and rules should be presented in user-understandable forms. As such, we organize the following summary of XAI advances 
on glioma imaging according to the aforementioned clinical-user-friendly explanatory forms.

Applying XAI methods to glioma imaging tasks is still at a nascent stage. The purposes of applying XAI in glioma imaging include model quality assessment, resolving users' perceived anomalies, gaining users' trust, improving physician expertise and assisting in scientific discovery (such as identifying imaging biomarkers).

\begin{enumerate}
    \item \textbf{Explaining using feature attributes}
    
Feature attribution is the most common form of explanation in the XAI literature. In an interview with clinicians about their requirements for XAI, doctors expressed the need to understand the important features that align with accepted medical paradigms~\cite{Tonekaboni2019}. Features are shown with their types of pathological changes, their importance scores, and their locations on the image. Such information can be presented in forms of text descriptions, color maps (or saliency maps) or segmentation maps overlaid unto input images. Next, we introduce three approaches to generate post-hoc explanations based on ``activation", ``gradient", and ``input perturbation".

\begin{enumerate}
    \item \textbf{Activation-based methods\\}
For deep learning-based image tasks using convolutional neural networks (CNNs), a usual approach to reveal what the network has learned is to utilize the information from the internal activation maps. For example, Ahmad et al. applied class activation map (CAM) method on 2D ResNet CNN to predict IDH mutation status in high-grade gliomas~\cite{Ahmad2019}. They demonstrated that the generated saliency maps correspond to the tumor area in most IDH1 mutant cases, while in most IDH1 wild-type cases, the peri-tumoral edema is also involved. The CAM method reveals where the intermediate CNN layer is ``looking at" to make the prediction. It has become popular in the computer vision literature since 2015~\cite{zhou2016learning}. 
Given an input image, it acquires the internal activation maps in a particular layer, and aggregates them using their respective weights from the global average pooling layer to generate a color map with the same size as the input image. Since the CAM method gives explanations on a single image, it is a local explanation method.

``Attention mechanism" is another approach to reveal the significant features in input data. It learns to assign ``importance weights" to activation maps (in the case of CNN) or hidden states (in the case of recurrent neural network). The weighted sum of the attention map can be overlaid on input data to show how the model attends to important regions in input data. For instance, Choi et al. built an attention-based long short-term memory (LSTM) neural network to identify IDH status from dynamic susceptibility contrast (DSC) perfusion MRI time-series data~\cite{Choi2019}. The model ``paid more attention" to the combination of the end of the pre-contrast baseline, up/downslopes of signal drops, and/or post-bolus plateaus for the curves used to predict IDH genotype.

\item \textbf{Gradient-based methods\\} 
Since gradient reflects the magnitude of output change in accordance with input change, it can provide users with clues of the important input features for the results. For example, to check if the model's predictions align with accepted medical knowledge, Meier et al. applied Grad-CAM and guided-backpropagation on 3D MRI CNN for classification of brain tumor grades~\cite{Meier2018}. By visually inspecting some of the saliency maps that do not contain the tumor regions, users may conclude that the prediction may not be reliable. 
Similarly, Huang et al. also applied 3D-based Grad-CAM on classifying high- vs. low-grade glioma to compare model performances on synthesized MR modalities with ground-truth complete modalities~\cite{Huang2019}. The resultant saliency map highlighted the tumor regions as the important image features that contribute to the tumor grading. 

Both grad-CAM and guided-backpropagation methods depend on the gradient of the output with respect to the input to compute 
how the change of input will influence the change of output.
Specifically, Grad-CAM, or gradient-weighted class activation map, can be regarded as a modification of the earlier CAM methods. Both utilize the weighted combination of activation maps to produce the saliency map; and the difference is in how they apply weighting to the activation maps. Unlike CAM that uses weights from global average pooling layer, grad-CAM takes weights from the gradients of the target class output with respect to the activation maps. Thus, it aggregates information both from activation maps and gradients, and can create a more localized saliency map than CAM~\cite{Selvaraju2016b}. Guided-backpropagation adds an additional guidance signal (gradient from top layer) to usual backpropagation which computes gradients from the bottom layer.
This can create clearer and less noisy saliency maps~\cite{Springenberg2014}.

\item \textbf{Perturbation-based methods\\} 
The above-mentioned methods apply to cases where we have access to model parameters, activations, and architecture information. Sometimes such knowledge is unknown to its model users, due to safety, privacy, or intellectual property issues. Perturbation-based method is used to reveal how the model makes predictions by probing the model with different inputs and analyzing the input-output pairs. It constitutes a post-hoc method and is model-agnostic (i.e., can be applied to any black-box models). In a work to segment gliomas from MRI, Pereira et al. applied the local interpretable model-agnostic explanations (LIME)~\cite{Ribeiro2016} to the ML model for local interpretability~\cite{Pereira2018a}. Their ML model is a restricted Boltzmann machine for unsupervised representation learning, and a random forest classifier for voxel-level segmentation. They perturbed the features to generate synthetic neighbours that are close to the original data in the feature representation space, and also acquired the predicted output of these neighbours. With the input-output pairs, they trained a ridge regressor and used its weights as the importance-measure of each feature, shown as saliency maps on the MRI images. The saliency maps showed that FLAIR sequences were the most important for segmenting the complete tumor while segmentation of edema was mainly based on FLAIR and T1c sequence. They also showed that the T2 and T1c sequences were important for segmenting necrosis, and T1c was important in segmenting the enhancing rim region. These findings correspond well with accepted radiologic principles; thus, confirming that the model learned the correct relations in the data. 
\end{enumerate}

\item \textbf{Explaining using examples\\}  
Compared to feature attribution, explaining using examples can provide more contextual information about how the model learned, and is more intuitive as clinicians are used to learning from examples. In a user study with ICU clinicians, the researchers observed the example-based explanations facilitated clinicians' analogical reasoning~\cite{Wang2019}. Another study involving pathologists showed that examples with similar visual appearance and different diagnoses can help physicians broaden their differential diagnosis~\cite{Cai}.

For example-based explanations applied to glioma imaging tasks, Chang et al. identified the prototypical images to predict IDH mutation, 1p/19q codeletion and MGMT methylation status~\cite{Chang1201}. They built CNN models on 2D MRI slices from the TCIA dataset and identified the prototypical images that caused the most activation of the units in the final fully-connected decision layer. The typical images for IDH mutation demonstrated absent or minimal enhancement with well-defined tumor margins in T1c, or central areas of cysts with FLAIR suppression. Conversely, IDH wild-type tumors exhibit thick and irregular enhancement or thin, irregular  peripheral enhancement in T1c, and infiltrative patterns of edema on FLAIR.
Such prototypical images serve as global explanations of the model's overall prediction. 
\end{enumerate}

\textbf{XAI evaluating metrics}
To evaluate interpretability methods, Doshi-Velez and Kim proposed an evaluation framework at three levels: functionally-grounded evaluation (using function as a proxy to measure interpretability), human-grounded evaluation (evaluating with lay-person's cognition and understanding), and domain expert-grounded evaluation (evaluating the interpretability on domain-specific tasks)~\cite{Doshi-Velez2017}.
Current XAI techniques on glioma imaging have little to no evaluation. This is probably due to the fact that the current XAI explorations are more focused on revealing the model's predictions and checking how well the learned model aligns with doctors’ prior knowledge, rather than considering the practical challenges associated with the clinical implementation of XAI. The research of clinical utility of XAI on glioma imaging is at its inception and has many unsolved questions waiting to be explored.

\section{Discussion}
In this manuscript, we have summarized recent efforts toward increasing the clinical utility of AI in neuro-oncology by addressing the current challenges in data collection, model training and clinical deployment. The current research, however, focuses more on novel algorithm design, rather than actually applying them in real-world patient-care settings. As the performance of machine and deep learning models improves and the challenges enumerated are addressed, we foresee greater interest in translational research that aims to apply novel technologies to the bedside.  To this end, we next highlight further advances which are necessary for the full implementation of AI in glioma imaging; to develop clinical decision support systems (CDSS) and facilitate the personalization of care.

\textbf{Clinical workflow integration}
Implementing AI in patient-care settings requires the AI technologies to be integrated into the existing clinical workflow. Several steps are needed to achieve clinical workflow integration: First, before the design of an AI system, clinical requirements should be ascertained and a ``needs assessment" should be performed. Furthermore, a user-centric design approach should be applied during the AI system design and iteration. This step solicits clinician  feedback in the design of user-friendly interfaces and reduces the cognitive-load for physicians while supporting their clinical tasks. Moreover, the implemented system needs to take advantage of newly acquired patient data using online machine learning, to continually improve the performance of the predictive models. The systems can also use ``adaptive learning" to adapt to user's behaviours and preferences, i.e. ``hybrid learning" or ``human-in-the-loop" machine learning~\cite{Holzinger2016}.
Achieving the above steps will need close collaboration between physicians, AI experts and human-computer interaction developers.

\textbf{Clinical evaluation}
Appropriately evaluating the functionality of AI systems is necessary prior to its clinical implementation. We can make an analogy of evaluating the clinical utility of AI to the conventional four-phase clinical trials for medications or medical devices: 

\begin{figure}[!ht]
\centering
\includegraphics[width=0.9\textwidth]{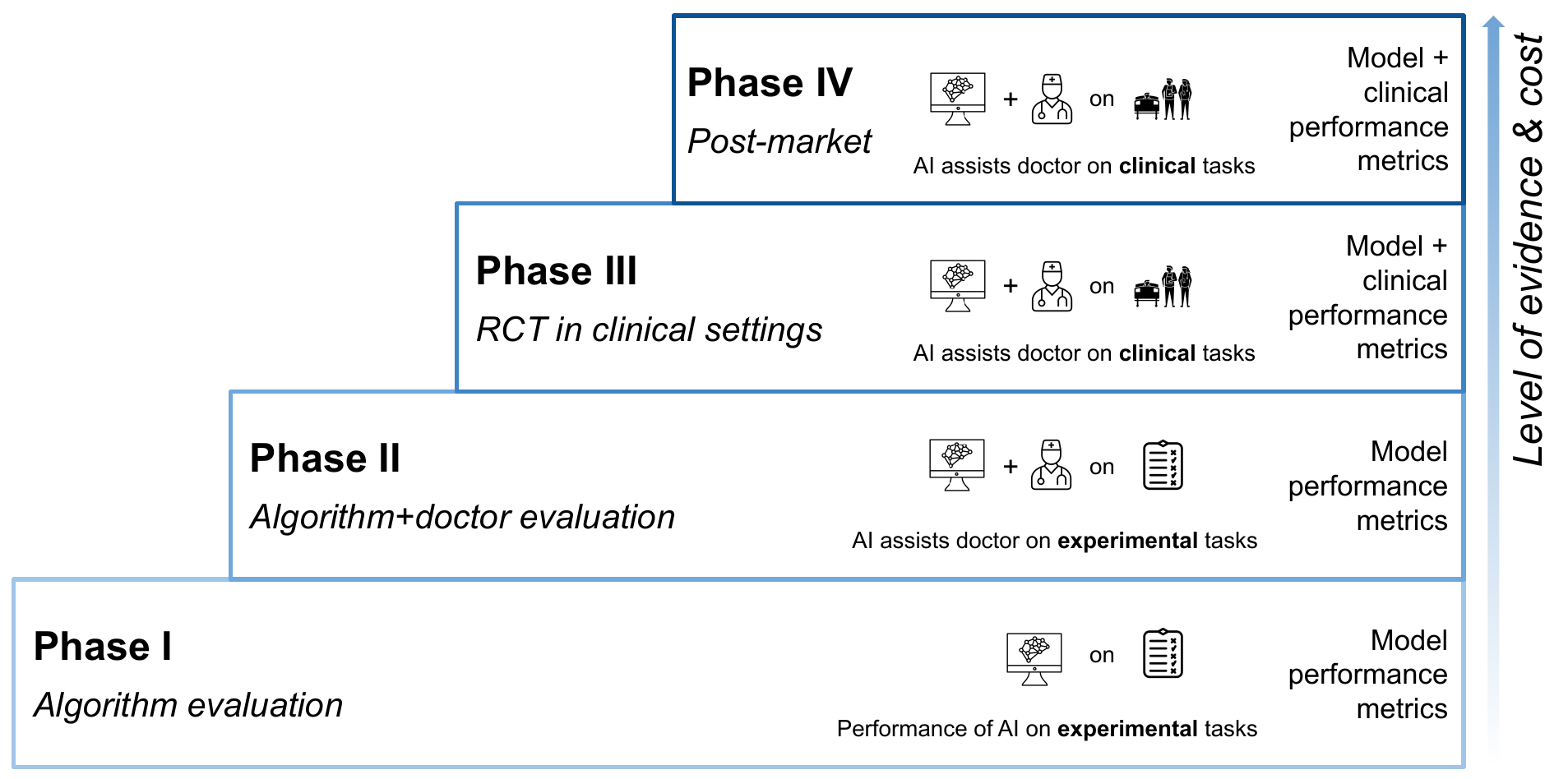}
\caption{The four phases of evaluating the clinical utility of AI in glioma imaging.}
\end{figure}

\begin{itemize}
    \item \textit{Phase I} is to primarily test the performance of the AI model. This phase only involves the AI model and test data, and does not involve human users. Most current AI system evaluation is at this stage and tests AI performance on unseen data with model-based evaluation metrics, such as classification accuracy, sensitivity, specificity, AUC (area under the receiver operating characteristics), or mean squared error, as listed in Table \ref{comparison}. For example, Li et al. created a clinical application called GliomaPredict that used an unsupervised ML principle component analysis (PCA) to classify patients with gliomas into 6 subtypes according to their glioma transcriptomic profiles~\cite{Li2010}. It generated a visual representation of the analyses, quantified the confidence of the underlying subtype assessment, and presented results as a printable PDF file. The application was evaluated based on the model performance on new patient data and yielded 75-96\% of prediction accuracy. 

    \item \textit{Phase II} involves clinical users in experimental settings using simulated tasks (usually on held-out test data) rather than in real clinical settings. The evaluation metrics are usually model-based metrics. Earlier works focused on the head-to-head comparison of doctors' performance with AI models, mainly to indicate that AI matches doctor-level performance~\cite{Esteva2017, Rajpurkar2017}. More recently, studies have involved doctor+AI as a third arm in the AI vs. doctor comparisons. For example, Bien et al. found that physicians who used AI predictions had significantly improved specificity in identifying anterior cruciate ligament tears from knee MRI compared to doctors who didn't have these predictions ~\cite{Bien2018}. This evaluation trend reflects a paradigm  shift from AI replacing doctors, to AI augmenting doctors.

    \item \textit{Phase III} involves clinical users in real-world settings using randomized controlled trials. In addition to using model-based evaluation metrics, other clinical outcome metrics can be evaluated, such as patients' outcome, physicians' performance or efficiency, clinical usability and health economics. So far, very few studies have entered this phase. In a randomized controlled trial, Wang et al. enrolled a total of 1,058 patient to evaluate an AI-assisted polyp detection system during colonoscopy~\cite{pmid30814121}. Compared with standard colonoscopy, the AI-assisted colonoscopy increased the adenoma detection rates by 50\%, from 20\% to 30\%.

    \item \textit{Phase IV} is for post-marketing software support and surveillance. This phase will follow the launch and clinical implementation of AI systems. 

\end{itemize}

\section{Conclusions}
Because of their significant clinical burden, gliomas are one of the most highly studied cancers. Machine and deep learning techniques have the potential to increase the clinical yield of non-invasive imaging studies, from segmenting tumors for quantitative measures, to identifying molecular subtypes and grades; from evaluating treatment response to predicting patient prognosis. Despite recent advances, AI technologies have not yet become fully functional in the management of patients with gliomas. In our review, we have summarized recent efforts to address practical challenges in applying AI to clinical settings. Multiple image and label collection efforts, together with data imputation and synthesis methods, were presented. We also discussed different model training strategies, and methods of learning under limited supervision.
Furthermore, we have presented various standardized model evaluation metrics and model interpretability methods.
Finally, we have proposed necessary future steps towards workflow integration and provided a framework for evaluating the clinical utility of AI-assisted systems. 
Due to the scope of the paper, we have focused on the technical barriers to implementation; and,  approaches to overcome these challenges. Specifically, we have not discussed non-technical, financial and legal issues which will, no doubt, need to be considered in the future. Ultimately, the full implementation of AI-based tools in neuro-oncology will significantly improve the care of patients with gliomas.

\ack
Partial funding for this project is provided by the Natural Sciences and Engineering Research Council of Canada (NSERC) and Simon Fraser University Big Data The Next Big Question Fund. 

\section*{References}
\raggedright
\bibliographystyle{unsrt}  
\bibliography{brain_review}
\end{document}